\documentstyle[preprint,eqsecnum,aps]{revtex}
\begin{document}
%%%%%%%%%%%%%%%% READ THIS %%%%%%%%%%%%%%%% READ THIS %%%%%%%%%%%%%%
%%%%%%%%%%%%%%%%%%%%%%%%%%%%%%%%%%%%%%%%%%%%%%%%%%%%%%%%%%%%%%%%%%%%%
% This paper has figures appended in  a second part as a uuencoded
% compressed tar file with instructions for unpacking.  They will be
%automatically
% included in the text if you have a functioning epsf.tex.
% If you don't have that macro package (available from hep-th), or don't
% have the figure files, COMMENT OUT THE FOLLOWING LINE:
\input epsf
% If you do not already have epsf.tex (it comes with the dvips driver),
% you can print out the postscript files separately.
% WARNING: there is more than one version of epsf.tex dated 18 Jul 1990
% some figures will produce errors unless you have the most recent
% version. note that the version of epsf.tex that comes with the
% NeXTstep 2.0/2.1 distribution is not an up-to-date version. you should
% get epsf.tex from hep-th if your dvips is up-to-date but your epsf.tex
% is not.
%
%%%%%%%%%%%%%%%%%%%%%%%%%%%%%%%%%%%%%%%%%%%%%%%%%%%%%%%%%%%%%%%%%%%%%%%%%%%
\tighten
\draft
\preprint{UCSBTH-97-04,NSF-ITP 97-024}

\title{Generalized Quantum Theory in\\ 
Evaporating Black Hole Spacetimes\thanks{Talk presented at {\sl Black
Holes and Relativistic Stars: A Symposium in Honor of S.~Chandrasekhar},
Chicago, IL, December 14--15, 1996.}}

\author{James B.
Hartle\thanks{e-mail: hartle@itp.ucsb.edu}}
\vskip .13 in
\address{Institute of Theoretical Physics\\
and Department of Physics,\\
University of California,\\
Santa Barbara, CA
93106-4030}

\maketitle

\date{\today}
\begin{abstract}
Quantum mechanics for matter fields moving in an evaporating black hole
spacetime is formulated in fully four-dimensional form according to the
principles of generalized quantum theory. The resulting quantum theory
cannot be expressed in a $3+1$ form in terms of a state evolving unitarily
or by reduction through a foliating family of spacelike surfaces.  That
is because evaporating black hole geometries cannot be foliated by a
non-singular family of spacelike surfaces.  A four-dimensional notion of
information is reviewed. Although complete information
may not be available on every spacelike  surface, information is not
lost in a spacetime sense in an evaporating black hole spacetime. 
Rather complete
information is distributed about the four-dimensional spacetime. Black
hole evaporation is thus not in conflict with the principles of quantum
mechanics when suitably generally stated.
\end{abstract}


\setcounter{footnote}{0}
\section{Introduction}
\label{sec:intro}

The early '80s were a memorable time to be at Chicago when 
Chandra was writing {\sl The Mathematical Theory of Black Holes}
\cite{Cha83}.  His method was that of an explorer in many ways ---
voyaging through the complex landscape of equations that the classical
theory of black holes presents --- discovering novel perspectives,
relationships, and hidden symmetries.  We discussed these many times in
long walks near the lake front in Hyde Park on often very cold Sunday
afternoons. It is therefore a special pleasure for me to contribute to
this commemoration of Chandra's work 

In the prologue to the {\sl Mathematical Theory}, Chandra sums up his
views on black holes in a sentence: ``The black holes of nature are the
most perfect macroscopic objects there are in the universe: the
only elements in their construction are our concepts of space and time.''
I note that Chandra used the word
``macroscopic'' to qualify the black holes that exhibit perfection. That
was either prudent or prescient, for now in theoretical physics we are
engaged with the question of whether black holes are or are not a blot
on the perfection of quantum theory --- the organizing principle of {\it
microscopic} physics. This essay puts forward the thesis that black
hole evaporation is not inconsistent with the principles of quantum
mechanics provided those principles are suitably generalized from their
usual flat space form.

\section{Black Holes and Quantum Theory}
\label{sec:II}
\subsection{Fixed Background Spacetimes}
\label{subsec:A}

The usual formulations of quantum theory rely on a fixed, globally
hyperbolic\footnote{A spacetime is {\it globally hyperbolic} if it
admits a surface $\Sigma$, no two points of which can be connected by a
timelike curve, but such that every inextendible timelike curve in the
spacetime intersects $\Sigma$.  Such a surface is
called a {\it Cauchy surface}. Classically data on a Cauchy surface
determine the entire future and past evolution of the spacetime.
Globally hyperbolic spacetimes have topology ${\bf R} \times \Sigma$ and
can be foliated by a one-parameter family of Cauchy surfaces defining a
notion of time. For more details see \cite{HE73}.}
 background
spacetime geometry as illustrated in Figure 1.
In these usual formulations, complete information about a physical
system is available on any spacelike (Cauchy) surface and is summarized
by a state vector associated with that surface.  This state vector
evolves through a foliating family of spacelike surfaces $\{\sigma\}$,
either
unitarily between spacelike surfaces
\begin{mathletters}
\label{twoone}
\begin{equation}
|\Psi\left(\sigma^{\prime\prime}\right)\rangle = U |\Psi
\left(\sigma^\prime\right)\rangle\ ,
\label{twoone a}
\end{equation}
or by reduction on them
\begin{equation}
|\Psi \left(\sigma\right)\rangle \to \frac{P|\Psi
\left(\sigma\right)\rangle}{||P|\Psi\left(\sigma\right)\rangle||}\ .
\label{twoone b}
\end{equation}
\end{mathletters}
\vskip .13in
\centerline{\epsfysize=3.00in \epsfbox{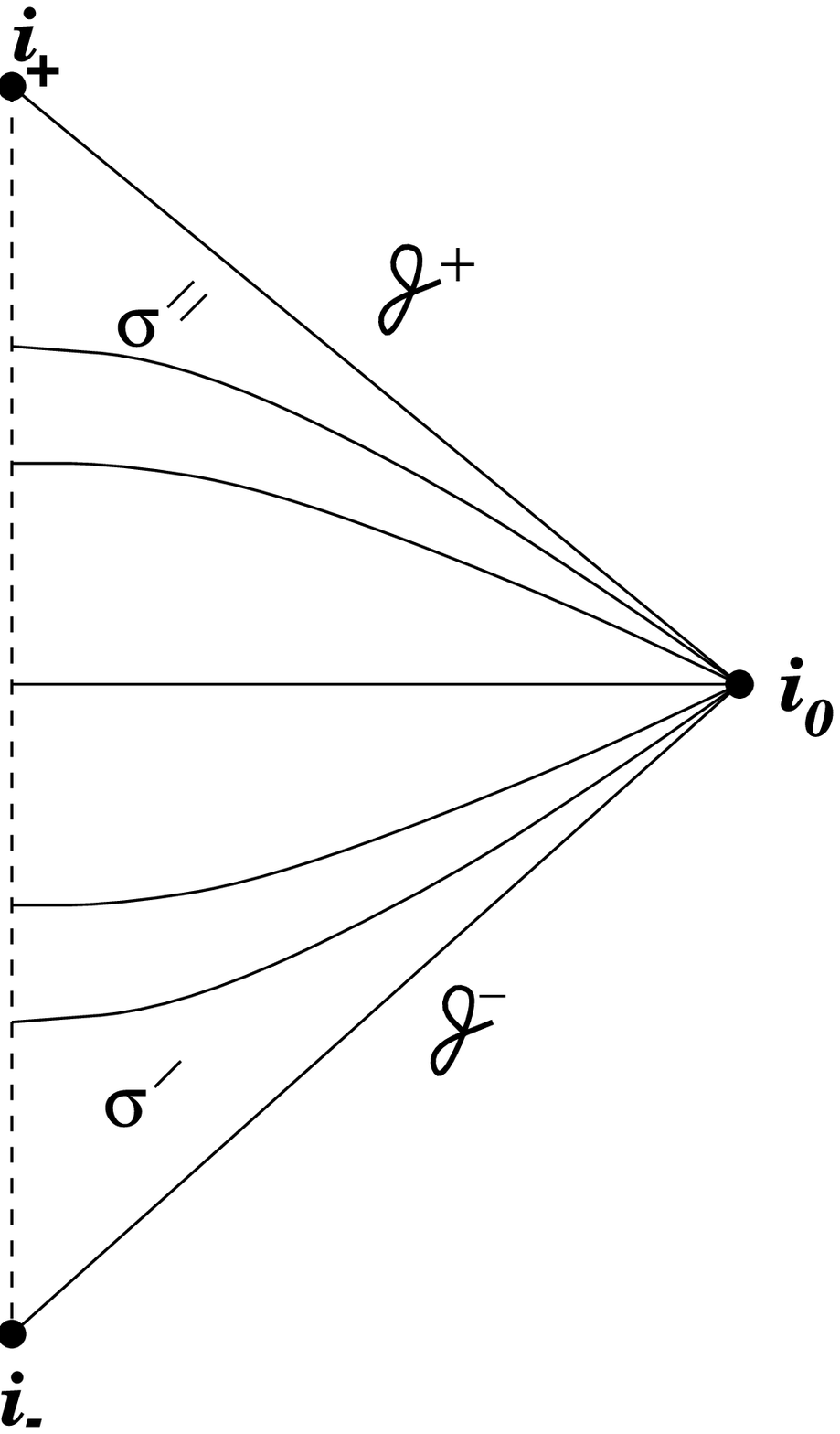}}
\vskip .13 in
\begin{quote}
{{\bf Figure 1}:  {\sl The Penrose diagram
for a globally hyperbolic spacetime like flat Minkowski
space.\footnote{\sl
 Spacetime has been rescaled so that it is contained in the interior of
the triangle and so that radial light rays move on $45^\circ$ lines.
The dotted line is radius $=0$. The remaining boundaries are the various
parts of infinity. Future null infinity, ${\cal I}^+$, consists of the
endpoints of light rays that escape to infinity. The points $i_0$, $i_+$,
and $i_-$ are respectively spacelike infinity --- where spacelike surfaces
that reach infinity end --- and future and past timelike infinity which are
the endpoints of timelike curves.  We shall use similar conventions in
other illustrations in this paper.  For more on the construction of
Penrose diagrams see \cite{HE73}.}
A globally hyperbolic spacetime may be smoothly foliated by a family of
spacelike surfaces a few of which are shown.  When a spacetime can be so
foliated quantum evolution can be described by a state vector that
evolves unitarily between surfaces or by reduction on them.}}
\end{quote}
Usual formulations of quantum theory thus depend crucially on
a background spacetime exhibiting non-singular foliations by spacelike
surfaces to define the notion of quantum state and its evolution.

Nowhere does the close connection between spacetime structure and 
quantum theory emerge so strikingly as in the process of black hole
evaporation. Black holes and quantum theory have been inextricably
linked since Hawking's 1974 \cite{Haw74} discovery of the tunneling
radiation from black holes that bears his name. That radiation when
analyzed in the approximation that its back reaction on the black hole
can be neglected requires nothing new of quantum theory.

The familiar story is summarized in the Penrose diagram in Figure 2. The
entire region of spacetime outside the horizon is foliable by a
non-singular family of spacelike surfaces.  States of matter fields
defined on a spacelike, pre-collapse surface $\sigma^\prime$ in the far
past evolve unitarily to states on later spacelike surfaces like
$\sigma^{\prime\prime}$. The complete information represented by the
state is available on any spacelike surface.  The evolution defines a
pure state $|\Psi(H,{\cal I}^+)\rangle$
\vskip .13 in
\centerline{\epsfysize=3.00in \epsfbox{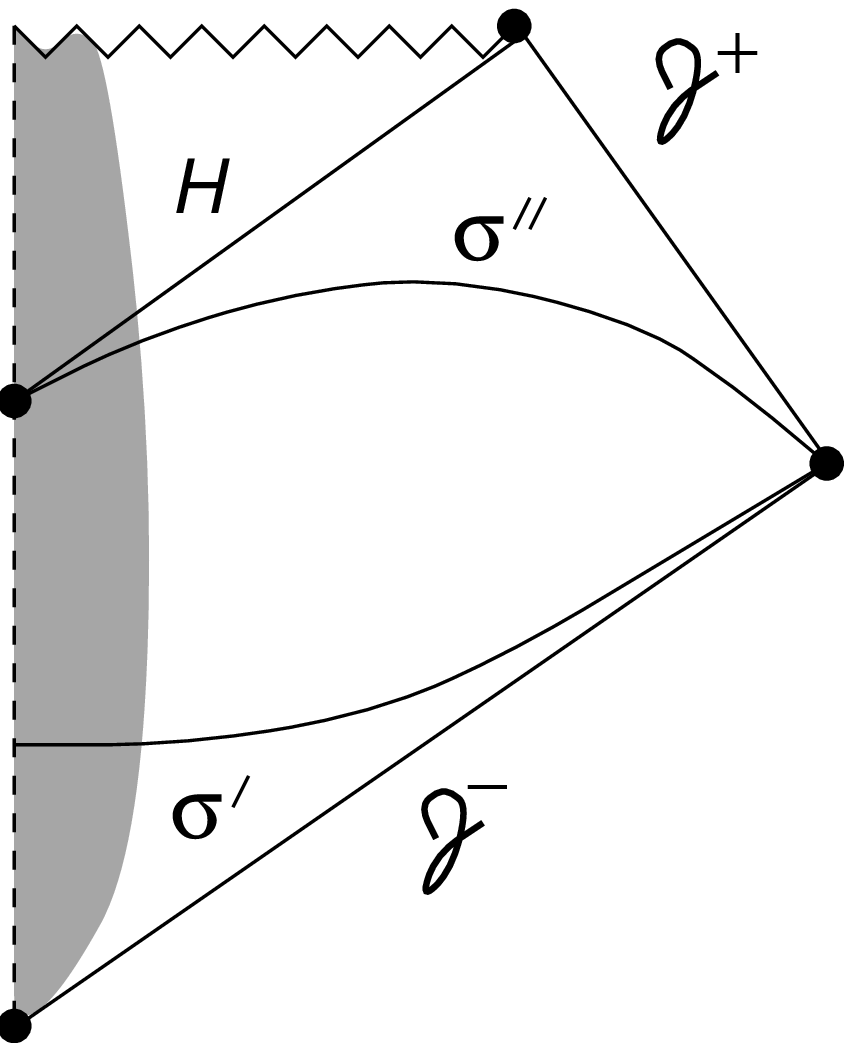}}
\vskip .13 in
\begin{quote}
{{\bf Figure 2:}  {\sl The Penrose diagram for a black hole when the back
reaction of the Hawking radiation is neglected. 
The shaded region in this diagram represents
spherically symmetric matter collapsing to form a black hole with
horizon $H$ and eventually a spacelike singularity represented
by the jagged horizontal line.  Light rays emitted from a point inside
the horizon end move on 45$^\circ$ lines, end 
at the singularity, and never escape to infinity.

A state of a 
field in this background
on an initial surface $\sigma^\prime$ can evolve unitarily through
an interpolating family of spacelike surfaces to a surface like
$\sigma^{\prime\prime}$. By pushing $\sigma^{\prime\prime}$ forward the
state can be evolved unitarily to the boundary of the region exterior to
the black hole consisting of ${\cal I}^+$ and the horizon $H$. 
The state $|\Psi({\cal I}^+,H)\rangle$ on ${\cal I}^+ \cup H$ exhibits
correlations (entanglements) between ${\cal I}^+$ and $H$.  Thus
complete information cannot be recovered far from the black hole on
${\cal I}^+$. Indeed for late times the predictions of $|\Psi({\cal I}_+,
H)\rangle$ for observations on ${\cal I}^+$ are indistinguishable from those of
a thermal density matrix at the Hawking temperature.  This does not mean
that information is lost in quantum evolution; it is fully recoverable
on the whole surface ${\cal I}^+ \cup H$.}}
\end{quote}
\vskip .13 in
on the surface $H\cup{\cal
I}^+$ consisting of the horizon and  future null-infinity. The Hilbert
space of states is the product ${\cal H}(H)\otimes {\cal H}({\cal I}^+)$
of field states on the horizon and on future null-infinity.
Measurements of observers at infinity probe only ${\cal H}({\cal I}^+)$.
The probabilities of their outcomes
may therefore be predicted from the density matrix
\begin{equation}
\rho({\cal I}^+) = tr_H \left[\bigr|\Psi(H,{\cal
I}^+)\rangle\langle\Psi(H,{\cal I}^+)\bigr|\right]
\label{twotwo}
\end{equation}
that results from tracing over all the degrees of freedom on the
horizon. For observations at late times this represents disordered,
thermal radiation.

Of course, information is missing from $\rho({\cal I}^+)$ so that complete
information is not available on ${\cal I}^+$.  Specifically, the
information in $|\Psi(H,{\cal I}^+)\rangle$ concerning correlations
between observables on $H$ and ${\cal I}^+$ is missing
\cite{Isr76,Wal75}. However, it is only necessary to consider observables
on \hfill\break
\vskip .13in
\centerline{\epsfysize=3.00in \epsfbox{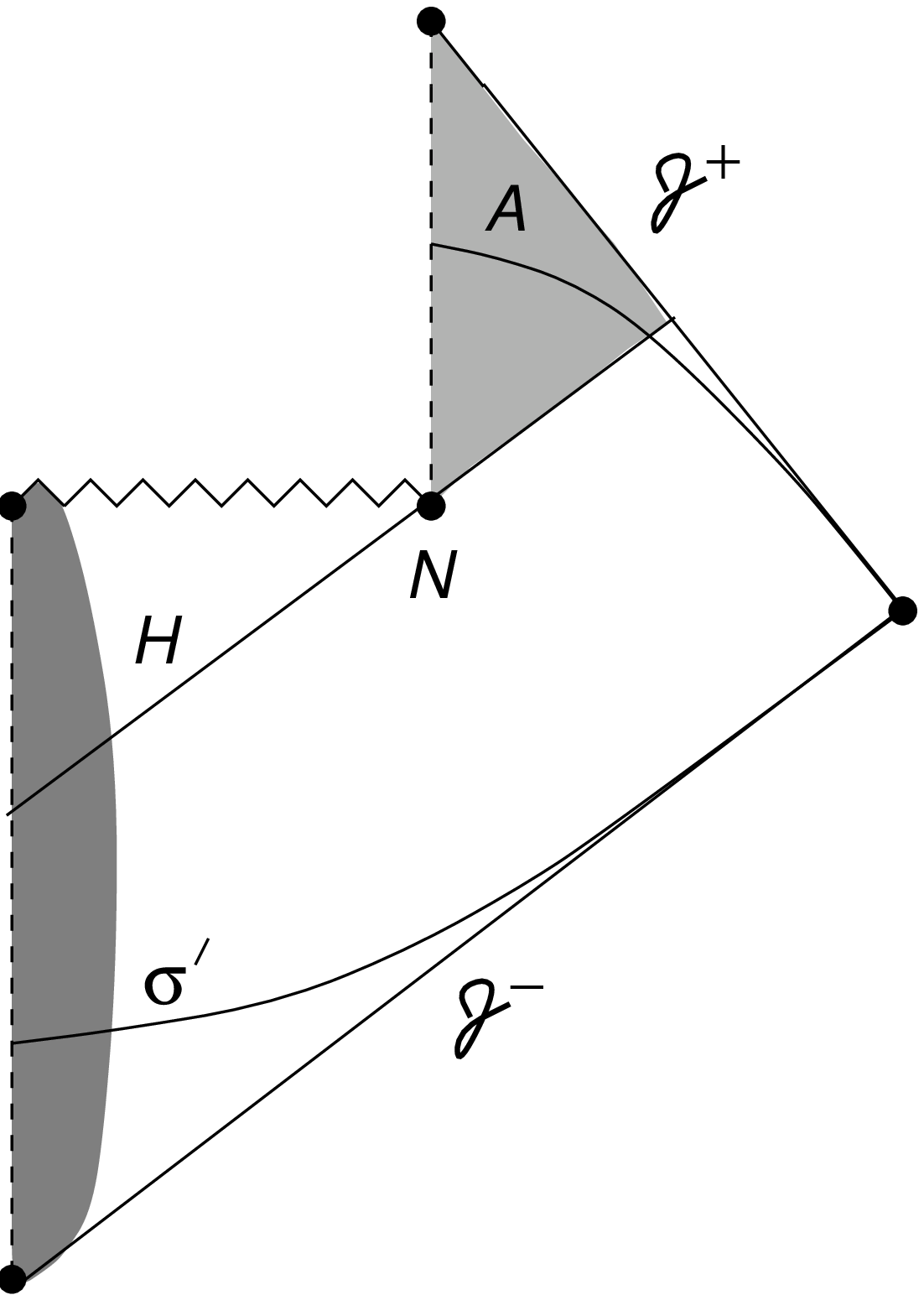}}
\vskip .13 in
\begin{quote}
{{\bf Figure 3:} {\sl The Penrose diagram for an evaporating black hole
spacetime.  The black hole is assumed here to evaporate completely, giving
rise to a naked singularity $N$, and 
leaving behind a nearly flat spacetime region (lightly shaded above).  
A spacelike
surface like $A$ is complete and data on this surface completely
determine the evolution of fields to its future. 
Yet complete information about a quantum matter field moving in
this spacetime is not available on $A$. Even if the initial state of the
matter field on a surface like $\sigma^\prime$ is pure, the state of the
disordered Hawking radiation on $A$ would be represented by a density matrix.  A
pure state cannot evolve unitarily into a density matrix, so the usual
formulation of quantum evolution in terms of states evolving through a
foliating family of spacelike surfaces breaks down.  
The geometry of evaporating
black hole spacetimes suggests why.  There is no smooth family of
spacelike surfaces interpolating between $\sigma^\prime$ and $A$ and
even classically there is not a well defined notion of evolution
of initial data on $\sigma'$ to $A$. The usual notion
of quantum evolution must therefore be generalized to apply in spacetime
geometries such as this.}}
\end{quote}
\vskip .13 in
both $H$ and ${\cal I}^+$ to recover it.  Usual quantum mechanics
with its notion of a state carrying complete information evolving
unitarily through families of spacelike surfaces is adequate to discuss
the Hawking radiation when its back reaction is neglected.

Only in the complete evaporation of a black hole does one find a hint
that the usual framework may need to be modified. Spacetime geometries
representing a process in which a black hole forms and evaporates
completely have a causal structure summarized by the Penrose diagram in
Figure 3.  Let us consider for a moment the problem of quantum mechanics
of matter fields in a fixed geometry with this causal structure. (We
shall return later to the
fluctuations in spacetime geometry that must occur
in a quantum theory of gravity\hfill\break
\vskip .13in
\centerline{\epsfysize=3.00in \epsfbox{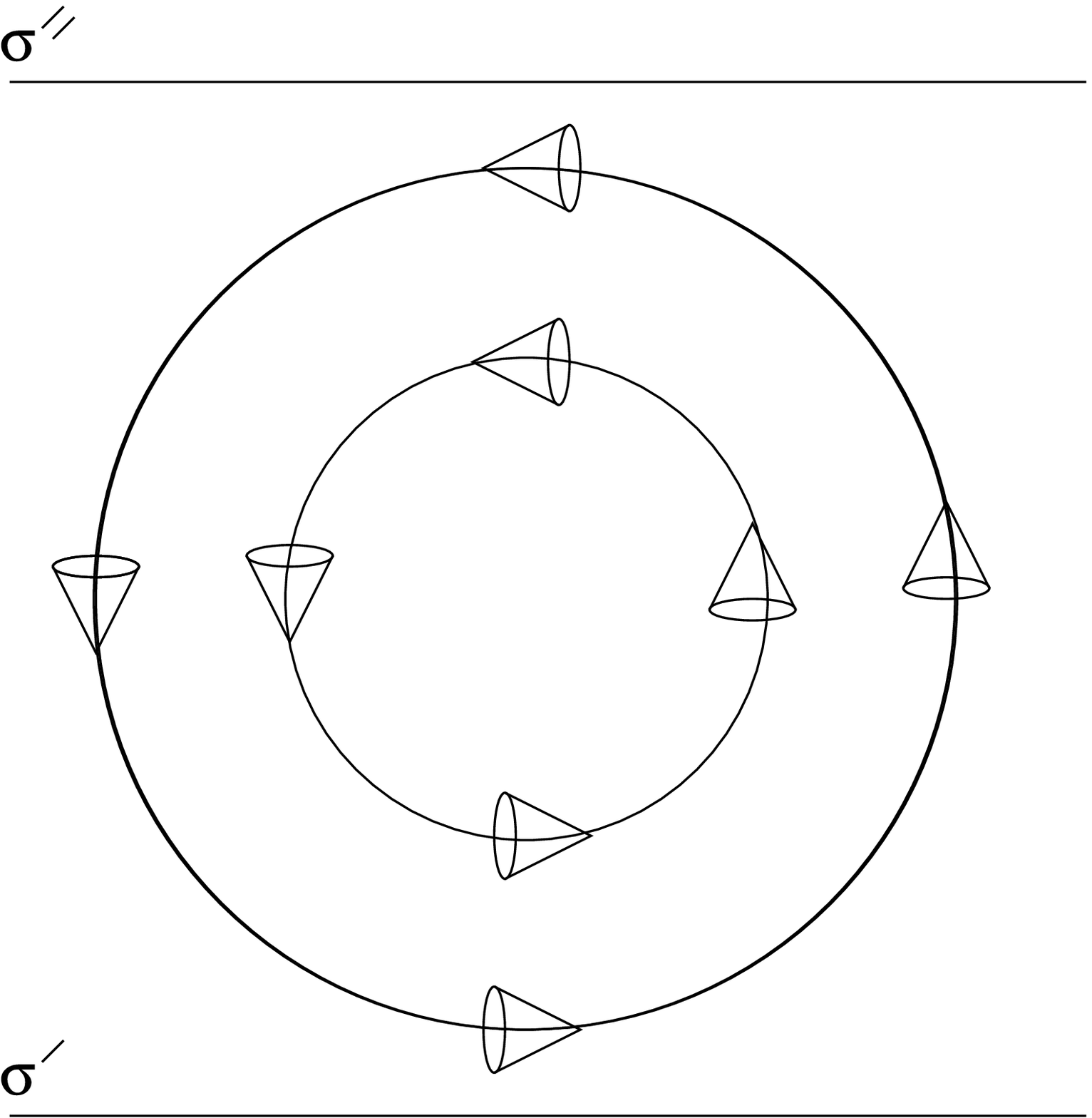}}
\vskip .13 in
\begin{quote}
{{\bf Figure 4:} {\sl 
A spacetime with a compact region of closed timelike curves (CTC's). 
As a consequence of the CTC's there is no family of
spacelike surfaces connecting an initial spacelike surface
$\sigma^\prime$ with a final one $\sigma^{\prime\prime}$, both outside
the CTC region.  
The quantum evolution of a matter field therefore cannot be described by a state
evolving through a foliating family of spacelike surfaces. The notion of
quantum evolution must be generalized to apply to spacetimes such as
this.}}
\end{quote} 
\vskip .13 in 
 that is  necessary to fully describe the
evaporation process.) Any pure initial state
$|\Psi(\sigma^\prime)\rangle$
leads to a state of disordered radiation on a hypersurface
$A$ after the black hole has evaporated,
so we  have
\begin{equation}
|\Psi(\sigma^\prime)\rangle \langle\Psi(\sigma')| \to \rho
\left(A \right)\ ,
\label{twothree}
\end{equation}
where $\rho(A)$ is the mixed density matrix describing the
radiation. This cannot be achieved by unitary evolution.  Indeed, it is
difficult to conceive of {\it any} law for the evolution of a
density matrix $\rho(\sigma)$ through a family of spacelike surfaces
that would
result in (\ref{twothree}) because there is no non-singular family of
spacelike surfaces that interpolate between $\sigma^\prime$ and
$A$. Even classically there is no well defined notion of evolution of initial
data on $\sigma'$ to the surface $A$ because of the naked singularity $N$.

I shall not attempt to review the discussion this situation has
provoked.\footnote{For reviews of this discussion see \cite{EVAPsum}.}  
What is clear is that a generalization of quantum mechanics
is needed for a quantum theory of fields in geometries such as this.

Evaporating black hole geometries are not the only backgrounds whose
causal structure requires a generalization of usual quantum mechanics
for a quantum theory of matter fields.
Consider a spacetime with a compact region of closed timelike
curves such as occur in certain \hfill\break
\vskip .13in
\centerline{\epsfysize=3.00in \epsfbox{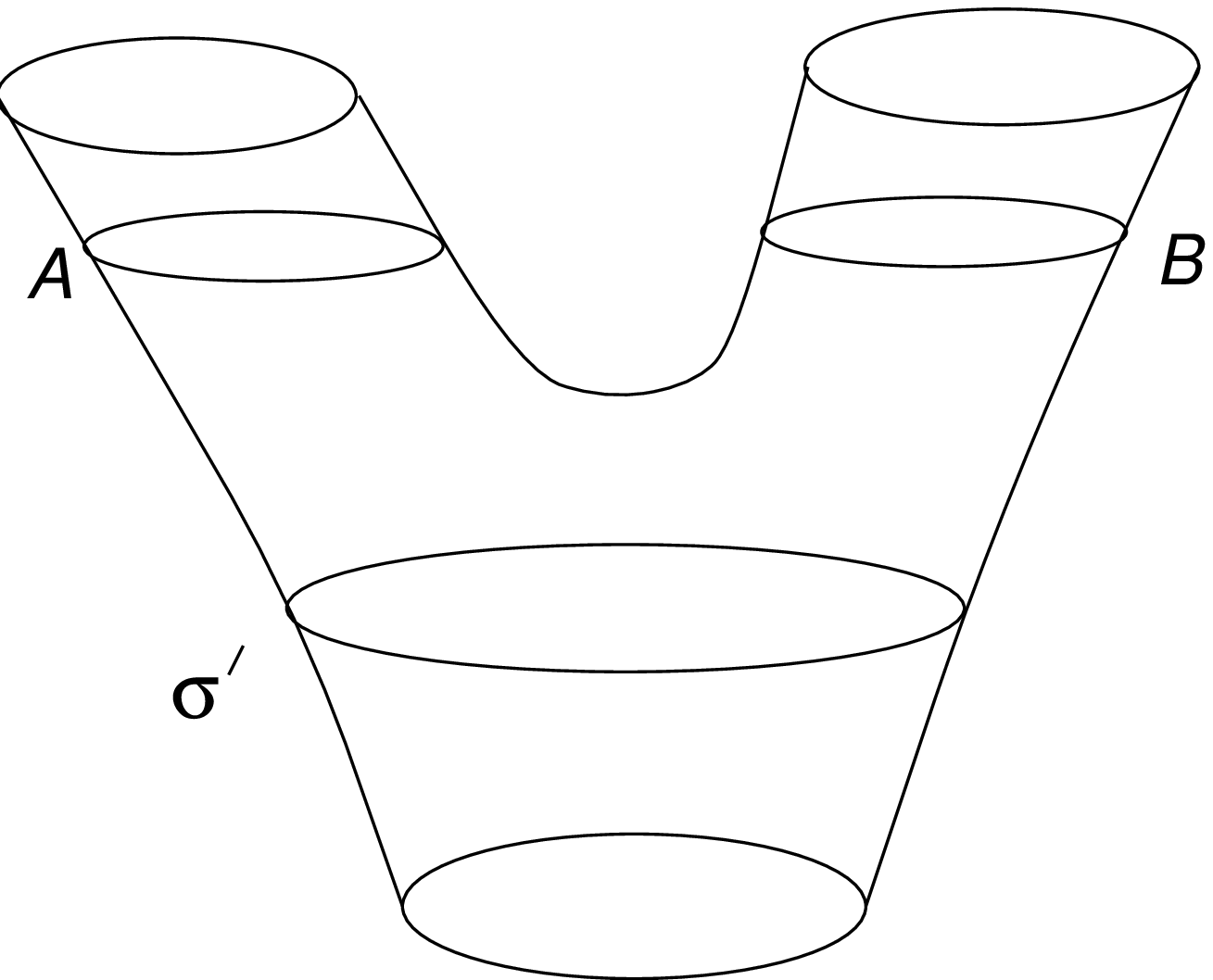}}
\vskip .13 in
\begin{quote}
{{\bf Figure 5:}
{\sl A spacetime with a simple change in spatial
topology.  There is no non-singular family of spacelike surfaces
between an ``initial''  spacelike surface $\sigma^\prime$ and ``final''
spacelike surfaces $A$ and $B$. The usual notion of quantum evolution
must therefore be generalized to apply to spacetimes such as this.
Complete information about the initial
state is plausibly not available on spacelike surfaces $A$ and
$B$ separately, but only surfaces $A$ and $B$ together.}}
\end{quote}
\vskip .13 in
\noindent 
 wormhole geometries
\cite{WORM}. (See Figure 4.) Given a state on a spacelike surface
$\sigma^\prime$ before the non-chronal region of closed timelike curves,
how do we calculate the
probabilities of field measurements inside the non-chronal region or
indeed on any spacelike surface like $\sigma''$ such that the non-chronal
region is contained between it and $\sigma'$? Certainly not
by evolving a state or density matrix through an interpolating family of
spacelike surfaces, whether by a unitary or non-unitary rule of
evolution.  
No such foliating family of spacelike surfaces exists! A
generalization of usual quantum mechanics is required.

Spacetimes exhibiting spatial topology change, as in the ``trousers''
spacetime of Figure 5, are another class of backgrounds for which
quantum field theory requires a generalization of usual quantum
mechanics. Given an initial state on a spacelike surface $\sigma^\prime$,
one could think of calculating the probabilities of alternatives on
spacelike surfaces $A$ or $B$. But because such spacetimes are necessarily
singular \cite{Ger67}, there is no smooth family of surfaces
interpolating between $\sigma^\prime$ and $A$ and $B$.  A generalization of
usual quantum dynamics is again required.\footnote{Field theory in such
spacetimes has been studied by a number of authors \cite{ADeW86,Hor91}.}
 
\subsection{Quantum Gravity}
\label{subsec:B}

The evaporating black hole spacetimes illustrated in Figure 3 are
singular. Spacetimes that are initially free from closed timelike curves
but evolve them later must be singular or
violate a positive energy condition
\cite{Tip77}. Spatial topology change implies either a singularity or
closed timelike curves \cite{Ger67}. These
pathologies suggest a breakdown in a purely classical description of
spacetime geometry.
One might therefore hope that the difficulties with usual 
formulations of quantum
theory in such backgrounds could be resolved in a quantum theory of
gravity.  The recent successful calculation of black hole entropy in
string theory \cite{SBHsum} raises several questions related to this
hope.  First, there is
the question of whether there are general principles mandating
a connection
between the entropy and the logarithm of a number of states in any
sensible quantum theory of gravity.  More important for the present
discussion, however, is the question of
whether these calculations mean that black
hole evaporation can be described within usual quantum mechanics in
string theory. It is possible that string theory will yield a unitary
$S$-matrix between asymptotic pre-collapse and evaporated states.
However, as with any quantum theory of gravity, the need for a
generalization of the idea of unitary evolution of states through
spacelike surfaces is only more acute than it is for field theory
in fixed non-globally-hyperbolic spacetimes for the following reason:

We have seen how the usual
quantum mechanics of fields with evolution defined through states on
spacelike surfaces relies heavily on a fixed, globally hyperbolic,
 background geometry 
to define those surfaces. But in any quantum
theory of gravity spacetime geometry is not fixed. Geometry is a quantum
variable --- generally fluctuating and without definite value. Quantum
dynamics cannot be defined by a state evolving in a given spacetime; no
spacetime is given.  A generalization of usual quantum mechanics is thus
needed. This need for generalization becomes even clearer if one accepts
the hint from string theory that spacetime geometry is not fundamental.

In the rest of this paper we shall discuss some generalizations of
quantum theory that are applicable to the process of black hole
evaporation.  We shall discuss these primarily for the case of field
theory in a fixed background evaporating black hole spacetime.  
There we
can hope to achieve a concreteness not yet available in quantum theories
of gravity.  However, there is every reason to believe 
that the principles of the
main generalization we shall describe are implementable in quantum
gravity as well \cite{Har95c}.

\section{The ${\bf \$}$-Matrix}
\label{sec:III}

Hawking \cite{Haw82} suggested one way that the principles of quantum
mechanics could be generalized to apply to an evaporating black hole.
For transitions between asymptotic states in the far past and far future
employ, not a unitary $S$-matrix mapping initial states to final states,
but rather a $\$$-matrix mapping initial density matrices to final density
matrices:
\begin{equation}
\rho^f = \$ \rho^i\ .
\label{threeone}
\end{equation}
That way an initial pure state could evolve into a mixed density matrix
as the evaporation scenario represented in Figure 3 suggests.

Hawking gave a specific prescription for calculating the $\$$-matrix: 
Use Euclidean sums over histories to calculate Euclidean
Green's functions for the metric and matter
fields. Continue these back to Lorentzian
signature in asymptotic regions
where the continuation is well defined because
the geometry is flat. Extract the \$-matrix elements from these Green's
functions as one would for the product of two $S$-matrices.  When the
topology of the Euclidean geometries is trivial, the resulting \$-matrix
factors into the product of two $S$-matrices and (\ref{threeone}) implies
that pure states evolve unitarily into pure states.  However, the \$-matrix
generally does not factor if the
topology the Euclidean geometries is non-trivial 
and then pure states can evolve into mixed states.
It remains an open question whether this prescription in fact yields a
\$-matrix.  It is not evident, in particular, whether the mapping that
results preserves the
positivity and trace of the density matrices on which it acts.

However constructed, a \$-matrix connects only asymptotic initial
and final density matrices. A full generalization of quantum theory
would predict the probabilities of non-asymptotic observables ``inside''
the spacetime.  To this end, various interpolating equations that will
evolve a density matrix through a family of spacelike surfaces have been
investigated \cite{BPSsum,UW95}. These typically have the form
\begin{equation}
i\hbar \frac{\partial\rho}{\partial t} = [H,\rho] + \left({\rm additional
\ terms} \atop {\rm linear\ is\ \rho}\right)\ ,
\label{threetwo}
\end{equation}
where additional terms give rise to non-unitary evolution.  Perhaps
these could be adjusted to yield the \$-matrix of Hawking's construction
when evolved between the far past and far future \cite{SWHpc}. 
Such equations display
serious problems with conservation of energy and charge when the
additional terms in (\ref{threetwo}) are local, although Unruh and Wald
\cite{UW95} have demonstrated that only a little non-locality is enough to
suppress this difficulty at energies below the Planck scale.

{From} the perspective of this paper, conservation or the lack of
it is not the main problem with such modifications of the quantum
evolutionary law.  Rather, it is that the ``$t$'' in the equations is
not defined.
As we have argued in Section II, we do not expect to have a
non-singular, foliating family of spacelike surfaces in evaporating 
black-hole spacetimes through which to evolve an equation like
(\ref{threetwo}). A generalization of quantum mechanics even beyond such
equations seems still to be required.

\section{Think Four-Dimensionally}

In the previous section we have argued that quantum theory needs to be
generalized to apply to physical situations such as black hole
evaporation in which quantum fluctuations in the geometry of spacetime
can be expected or situations such as field theory in evaporating black
hole backgrounds where geometry is fixed, but not  globally
hyperbolic. What features of usual quantum mechanics
must be given up in order to achieve this generalization?  In this
section, we argue that one feature to be jettisoned is the notion of a
state on spacelike surface and quantum evolution described in terms of
the change in such states from one spacelike surface to another.

The basic argument for giving up on evolution by states through a
foliating family of spacelike surfaces in spacetime has already been
given:  When geometry is not fixed, or even when fixed but without an
appropriate causal structure, a foliating family of spacelike surfaces
is not available to define states and their evolution.
\vskip .13 in
\centerline{\epsfysize=3.00in \epsfbox{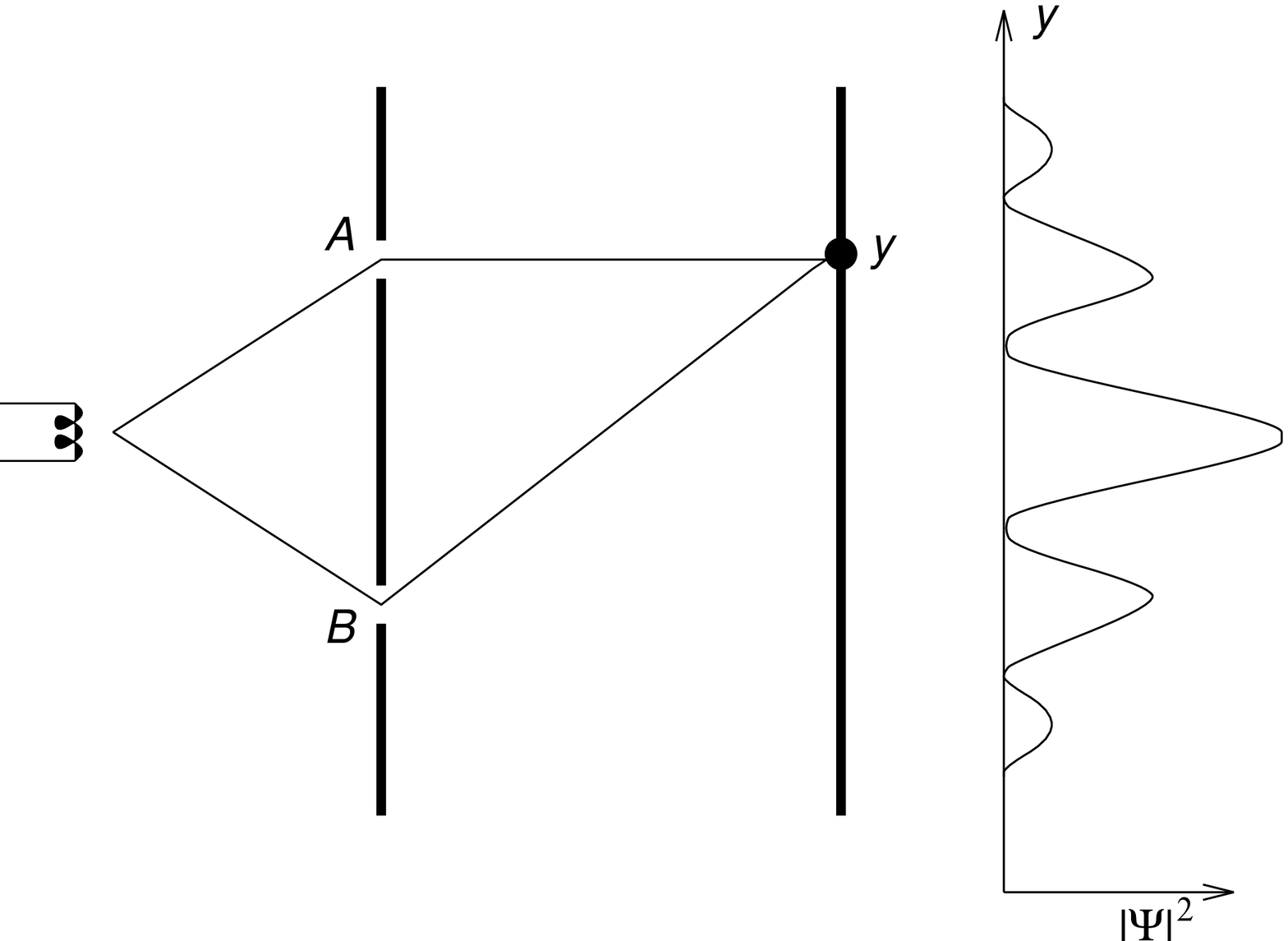}}
\vskip .13 in
\begin{quote}
{{\bf Figure 6:} The Two-Slit Experiment.
{\sl An electron gun at left emits an electron traveling towards a screen
with two slits and then on to detection at a further screen. Two
histories are possible for electrons detected at a particular point on
the right-hand screen defined by whether they went through slit A or
slit B. Probabilities cannot be consistently assigned to this set of two
alternative histories because they interfere quantum mechanically.}}
\end{quote}
\vskip .13 in
This basic point is already evident classically.  There is a fully 
four-dimensional description of any spacetime geometry in terms of
four-dimensional manifold, metric, and field configurations.  However,
for globally hyperbolic geometries that
four-dimensional information can be compressed into initial data on a
spacelike surface. That initial data is the classical notion of state.
By writing the Einstein equation in $3+1$ form the four dimensional
description  can be recovered by evolving the state through a
family of spacelike surfaces. However such compression is not possible
in spacetimes like the evaporating black hole spacetime illustrated in
Figure 3. Only a four-dimensional description is possible.

Similarly there is a fully four-dimensional formulation of quantum field theory
in background spacetime geometries in terms of Feynman's
sum-over-field-histories.  Transition amplitudes between spacelike
surfaces are specified directly from the four-dimensional action $S$ by
sums over field histories of the form
\begin{equation}
\sum\limits_{\rm histories} \exp[iS({\rm history})/\hbar]\ .
\label{fourone}
\end{equation}
When the background geometry is globally hyperbolic, these transition
amplitudes between spacelike surfaces can be equivalently calculated by
evolving a quantum state through an interpolating family of spacelike surfaces.
However, if the geometry is not globally  hyperbolic, 
we cannot expect such a $3+1$ formulation of
quantum dynamics any more than we can for the classical theory.
Following
Feynman, however, we expect fully four-dimensional spacetime formulation
of quantum theory to supply the necessary generalization\footnote{There
are other ideas for the necessary generalization. Wald has reached
similar conclusions in the algebraic approach to field theory in curved
spacetime \cite{Wal94}. The most developed is
the idea from canonical quantum gravity that states 
evolve, not through surfaces in spacetime, but rather through surfaces
in the superspace of possible three-dimensional geometries.  For lucid
and, by now, classic reviews of the various aspects and difficulties
with this approach see \cite{POT}.
These ideas are not obviously applicable to the cases of fixed spacetime
geometry discussed in this paper.  For another that might be, see
\cite{Deu91} where states are defined in local
regions later patched together.} applicable to field theory in
evaporating black hole spacetimes and to the other examples we have
mentioned.  We shall describe this
generalization and its consequences in this article.\footnote{The
application of sum-over-histories methods to gravity has a considerable
history that cannot be recapitulated here.  Some of the more notable
early references are \cite{SOHQG}.}
Our motto is: ``Think four-dimensionally.''
When we do there is no necessary conflict between
quantum mechanics and black hole evaporation.

\section{Generalized Quantum Theory}

To generalize usual quantum theory it is just as necessary to decide
which features of the usual framework to retain as it is to decide which
to discard.  Generalized quantum theory is a comprehensive
framework incorporating the essential features of a broad class
of generalizations of the usual theory.\footnote{For 
expositions see, {\it e.g.}~\cite{Ish94,Har95c}.}  The full apparatus of generalized
quantum theory is not essential to reach the conclusion that black-hole
evaporation is consistent with the principles of quantum mechanics
sufficiently generally stated.  Indeed, similar conclusions have been
reached without invoking these general principles. (See,
{\it e.g.}~\cite{Wal94}.) What is necessary is a generalization of the usual
notion of unitary evolution.  However, generalized quantum theory is a useful
setting to consider generalizations 
because it provides basic principles and
a framework to compare to compare different  generalizations. In this section
we give a brief and informal exposition of these principles.

The most general objective of a quantum theory is to predict the
probabilities of the individual members of a set of 
alternative histories of a closed system, most generally the universe.
The set of alternative orbits of the earth around the sun 
is an example.  These are sequences of positions of the
earth's center of mass at a series of times. Another example are the
four-dimensional histories of matter fields in an evaporating black hole
spacetime.

The characteristic feature of a quantum theory is that not
every set of alternative histories can be consistently assigned
probabilities because of quantum mechanical interference.  This is
clearly illustrated in the famous two-slit experiment shown in Figure 6.  
There are two
possible histories for an electron which proceeds from the source to a
point $y$ at the detecting screen. They are
 defined by which of the two slits (A
or B) it passes through.  It is not possible to assign
probabilities to these two histories.  It would be inconsistent to do so
because the probability to arrive at $y$ is not the sum of the
probabilities to arrive at $y$ {\it via} the two possible histories:
\begin{equation}
p(y) \not= p_A(y) + p_B(y)\ .
\label{fiveone}
\end{equation}
That is because in quantum mechanics probabilities are squares of
amplitudes and
\begin{equation}
\left|\psi_A(y) + \psi_B(y)\right|^2 \not= \left|\psi_A(y)\right|^2 +
\left|\psi_B(y)\right|^2
\label{fivetwo}
\end{equation}
when there is interference.

In a quantum theory a rule is needed to specify which sets
of alternative histories may be assigned probabilities and which may
not.  The rule in usual quantum mechanics is that probabilities can be
assigned to the histories of the outcomes of {\it measurements} and not
in general otherwise.  Interference between histories is destroyed by
the measurement process, and probabilities may be consistently assigned.
However, this rule is too special to apply in the most general
situations and certainly insufficiently general for cosmology.
Measurements and observers, for example, were not present in the early
universe when we would like to assign probabilities to histories of
density fluctuations in matter fields or the evolution of spacetime
geometry.

The quantum mechanics of closed systems\footnote{See, {\it e.g.}~\cite{Har93a}
for an elementary review and references to earlier literature.}
 relies on a more general rule
whose essential idea is easily stated: A closed system is in some
initial quantum state $|\Psi\rangle$. Probabilities can be assigned to
just those sets of histories for which there is vanishing interference
between individual histories
as a consequence of the state $|\Psi\rangle$ the system is in.
Such sets of histories are said to {\it decohere}. Histories of
measurements decohere as a consequence of the interaction between the
apparatus and measured subsystem. Decoherence thus contains the rule of
usual quantum mechanics as a special case.  But decoherence is more
general.  It permits assignment of probabilities to alternative orbits
of the moon or alternative histories of density fluctuations in the
early universe when the initial state is such that these alternatives
decohere whether or not the moon or the density fluctuations
are receiving the attention of observers
making measurements.

The central element in a quantum theory based on this rule is the
measure of interference between the individual histories
$c_\alpha, \ \alpha=1,2,\cdots$ in a set of alternative histories.  This measure is
called the {\it decoherence} {\it functional},
$D(\alpha^\prime,\alpha)$. A set of histories decoheres when
$D(\alpha^\prime,\alpha) \approx0$ for all pairs
$(\alpha^\prime,\alpha)$ of distinct histories.

The decoherence functional for usual quantum mechanics is defined as
follows: A set of alternative histories can be specified by giving
sequences of sets of yes/no alternatives at a series of times
$t_1,\cdots,t_n$. For example, alternative orbits may be specified by
saying whether a particle is or is not in certain position ranges at a
series of times.  Each yes/no alternative in an exhaustive set of
exclusive alternatives at time $t_k$ is represented by a Heisenberg picture
projection operator $P^k_{\alpha_k}(t_k)$, $\alpha_k = 1,2,\cdots$ where $\alpha$ labels
the different alternatives in the set. An 
individual history $\alpha$ is a particular
sequence of alternatives $\alpha \equiv (\alpha_1,\cdots,\alpha_n)$ and
is represented by the corresponding chain of Heisenberg picture
projections:
\begin{equation}
C_\alpha = P^n_{\alpha_n}(t_n) \cdots P^1_{\alpha_1}(t_1)\ .
\label{fivethree}
\end{equation}
If the initial state vector of the closed system is $|\Psi\rangle$,
(non-normalized) branch state vectors corresponding to the
individual histories $\alpha$ may be defined by
\begin{equation}
C_\alpha |\Psi\rangle\ ,
\label{fivefour}
\end{equation}
and the decoherence functional for usual quantum mechanics is:
\begin{equation}
D(\alpha^\prime,\alpha)= \langle\Psi
|C^\dagger_\alpha C_{\alpha^\prime}
|\Psi\rangle\ .
\label{fivefive}
\end{equation}
This is the usual measure of interference between different histories
represented in the form (\ref{fivefour}).

The essential properties of a decoherence functional that are necessary
for quantum mechanics may be characterized abstractly \cite{Ish94} and
(\ref{fivefive}) is only one of many other ways of satisfying these
properties.  Therein lie the possibilities for generalizations of usual
quantum mechanics. 

\section{Spacetime Generalized Quantum Mechanics}

Feynman's sum-over-histories ideas may be used with the concepts of
generalized quantum theory to construct a fully four-dimensional
formulation of the quantum mechanics of a matter
field $\phi(x)$ in a fixed background spacetime.
 We shall sketch this spacetime quantum
mechanics in what follows. We take the dynamics of the field to be
summarized by an action $S[\phi(x)]$ and denote the initial state of the
closed field system by $|\Psi\rangle$.

The basic (fine-grained) histories are the alternative, four-dimensional
field configurations on the spacetime. These may be restricted to
satisfy physically appropriate conditions at infinity and at the
singularities.
Sets of alternative (course-grained) histories to which the theory
assigns probabilities if decoherent are partitions of these field
configurations into exclusive classes $\{c_\alpha\}, \alpha=1,2,\cdots$.
For example, the alternative that the field configuration on a spacelike
surface $\sigma$ has the value $\chi({\bf x})$ corresponds to the class
of four-dimensional field configurations which take this value on
$\sigma$. In flat spacetime the probability of this alternative is the
probability for the initial state $|\Psi\rangle$ to
evolve to a state $|\chi ({\bf x}), \sigma\rangle$ of definite field on
$\sigma$. The history where the field takes the value $\chi^\prime({\bf
x})$ on surface $\sigma^\prime$ and $\chi^{\prime\prime}({\bf x})$ on a later
surface $\sigma^{\prime\prime}$ corresponds to the class of
four-dimensional field configurations which take these values on
the respective surfaces, and so on.

The examples we have just given correspond to the usual quantum
mechanical notion of alternatives at a definite moment of time or a
sequence of such moments. However,
more general partitions of four-dimensional field configurations are
possible which are not at any definite moment of time or series of such
moments.  For example, the four-dimensional field configurations could
be partitioned by ranges of values of their averages over a region
extending over both space and time.  Partition of four-dimensional histories
into exclusive classes is thus a fully four-dimensional notion of alternative
for quantum mechanics.

Branch state vectors corresponding to individual classes $c_\alpha$
in a partition of the fine-grained
field configurations $\phi(x)$ can be constructed from the sum over
fields in the class $c_\alpha$.  We write schematically
\begin{equation}
C_\alpha|\Psi\rangle = \int_{c_\alpha}
\delta\phi\exp\Bigl(iS[\phi(x)]/\hbar\Bigr)|\Psi\rangle\ .
\label{sixone}
\end{equation}
\vskip .26 in
\centerline{
\begin{tabular}{|l|l|l|}\hline
& Usual & Generalized \\
& Quantum Mechanics & Quantum Mechanics \\\hline\hline
Dynamics & $e^{-iHt}|\Psi\rangle$
& $\sum\limits_{{\rm histories}\atop
\in c_\alpha} e^{iS({\rm
history})/\hbar}|\Psi\rangle$ \\
&$P|\Psi\rangle/\Vert P|\Psi\rangle\Vert$&\\\hline
Alternatives & On spacelike surfaces or & Arbitrary partitions of\\
& sequences of surfaces & fine-grained histories\\\hline
Probabilities & Histories of measurement & Decohering sets\\
assigned to & outcomes & of histories\\\hline
\end{tabular}
}
\vskip .26 in
It is fair to say that the definition of such integrals has been  little
studied in interesting background spacetimes, but we proceed assuming a
careful definition can be given even in singular spacetimes such as those
discussed in Section II.  Even then some discussion is needed to
explain what (\ref{sixone}) means as a formal expression. In a globally
hyperbolic spacetime we can define an operator $C_\alpha$ 
corresponding to the
class of histories $c_\alpha$ by specifying the matrix elements
\begin{equation}
\left\langle\chi^{\prime\prime}({\bf x}),
\sigma^{\prime\prime}\left|C_\alpha\right| \chi^\prime({\bf x}),
\sigma^\prime\right\rangle = \int_{[\chi^\prime c_\alpha
\chi^{\prime\prime}]} \delta\phi\exp\Bigl(iS[\phi(x)]/\hbar\Bigr)\ .
\label{sixtwo}
\end{equation}
\noindent The sum is over all fields in the class $c_\alpha$ that match
$\chi^\prime({\bf x})$ and $\chi^{\prime\prime}({\bf x})$ on the
surfaces $\sigma^\prime$ and $\sigma^{\prime\prime}$ respectively.  This
operator can act on $|\Psi\rangle$ by taking the inner product with its
field
representative $\langle\chi^\prime ({\bf x}),
\sigma^\prime|\Psi\rangle$ on a spacelike surface far in the past.  By
pushing $\sigma^{\prime\prime}$ forward to late times we arrive at the
definition of $C_\alpha|\Psi\rangle$. The same procedure could be used
to define branch state vectors in spacetimes with closed timelike curves
(Figure 4), in spacetimes with spatial topology change (Figure 5), and
in evaporating black hole spacetimes (Figure 3). The
only novelty in the latter two  cases is that $C_\alpha|\Psi\rangle$ lives on
the product of two Hilbert spaces. There are the 
Hilbert spaces on the two legs of
the trousers in the spatial topology change case and, in the black hole
case, there are the Hilbert space of states inside
the horizon and the Hilbert space of states on late time surfaces after
the black hole has evaporated.

The decoherence functional is then
\begin{equation}
D(\alpha^\prime,\alpha) = {\cal N}\langle\Psi|C^\dagger_\alpha
C_{\alpha^\prime}|\Psi\rangle\ ,
\label{sixthree}
\end{equation}
where $\cal N$ is a constant to ensure the normalization  condition
\begin{equation}
\Sigma_{\alpha \alpha'} D(\alpha', \alpha) =1  \ .
\label{new}
\end{equation}

A set of alternative histories decoheres when the
off-diagonal elements of $D(\alpha^\prime,\alpha)$ are negligible. The
probabilities of the individual histories are
\begin{equation}
p(\alpha) = D(\alpha, \alpha) = {\cal N}\left\Vert C_\alpha
|\Psi\rangle\right\Vert^2\ .
\label{sixfour}
\end{equation}
There is no issue of ``conservation of probability'' for these
$p(\alpha)$; they are not defined in terms of an evolving state vector.
As a consequence of decoherence, the $p(\alpha)$ defined by
(\ref{sixfour}) obey the most general probability sum rules including,
for instance, the elementary normalization condition $\sum_\alpha p(\alpha)=1$
which follows from (\ref{new}). 
\vskip .13in
\centerline{\epsfysize=3.00in \epsfbox{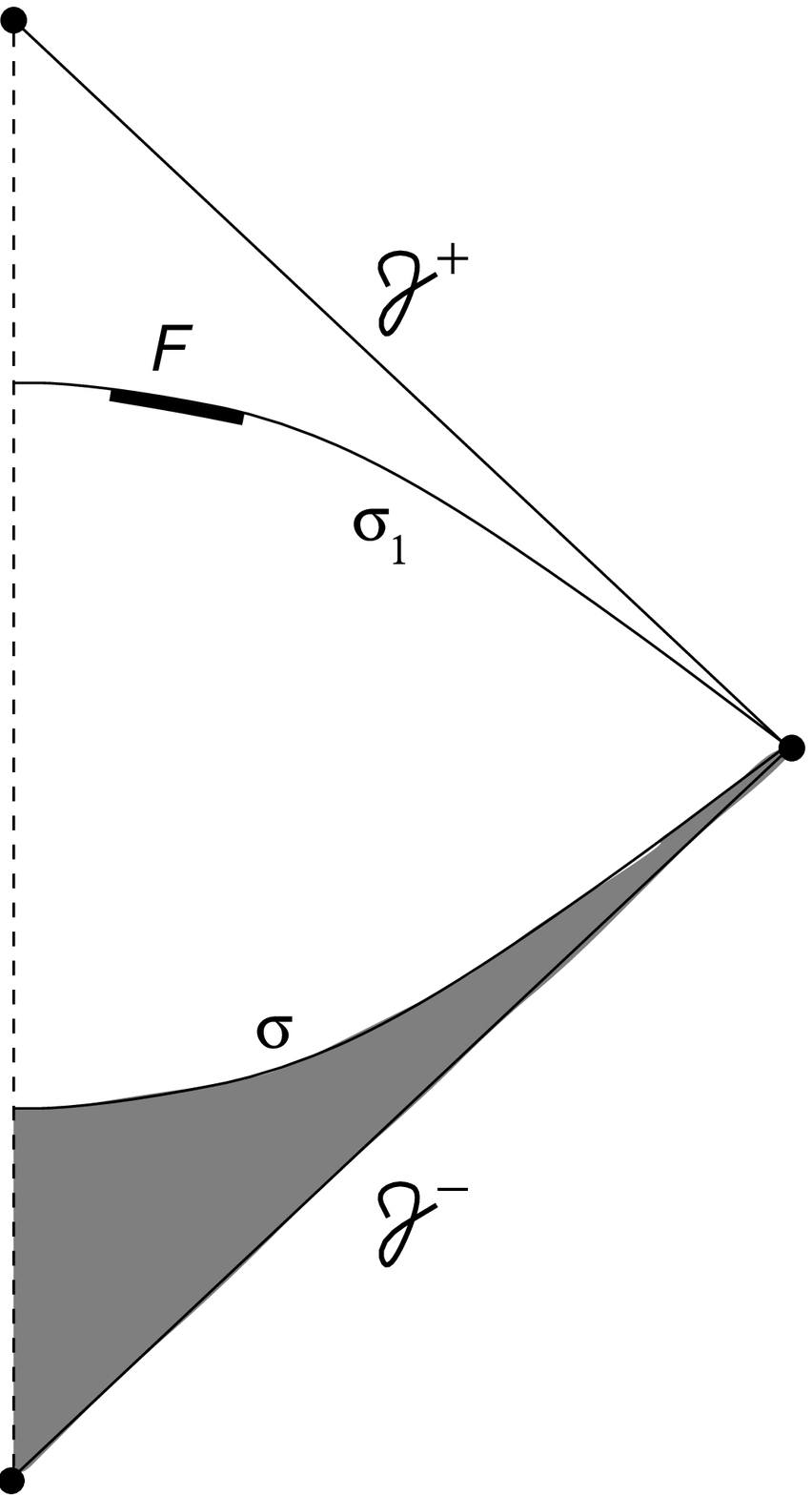}}
\vskip .13 in
\begin{quote}
{{\bf Figure 7:} {\sl  Recovery of the unitary evolution of states
through
spacelike surfaces from sum-over-histories quantum mechanics in a
globally hyperbolic spacetime. A sum-over-field-configurations
between ${\cal I}^-$ and a spacelike surface $\sigma$ defines a state on
that spacelike surface. This
sum-over-fields specifies the evolution of this state as $\sigma$ is
pushed forward in time.  In a globally
hyperbolic spacetime this evolution is unitary until a surface like
$\sigma_1$ is encountered where there is a restriction on the spatial
field
configurations to a class $F$.  This restriction is equivalent to the
action of a projection on the state.  Thus the two standard laws of
quantum evolution in a ``$3+1$'' formulation are recovered from the more
general four-dimensional framework. That  recovery, however, requires a
spacetime geometry that is smoothly foliable by spacelike surfaces.}}
\end{quote}
\vskip .13 in
This spacetime generalized quantum theory is only a modest
generalization of usual quantum mechanics in globally hyperbolic
backgrounds as the above table shows.  The two laws of evolution
(\ref{twoone}) have been unified in a single sum-over-histories
expression. The alternatives potentially assigned probabilities have
been generalized to include ones that extend in time and are not simply
the outcomes of a measurement process.  These generalizations 
put the theory in fully four-dimensional form.

When the more general framework is restricted to globally hyperbolic
backgrounds, to histories of alternatives on spacelike surfaces, and
to the outcomes of measurements, usual quantum theory is recovered as a special
case.  In particular, one recovers the notion of a state evolving either
unitarily or by reduction through a foliating family of spacelike
surfaces.  To see how this works consider the globally hyperbolic 
spacetime shown in Figure 7 and  a single
alternative where the field is restricted to some set $F$ of spatial
field configurations on a spacelike surface $\sigma_1$. The class
operator $C_\alpha$ is defined through
(\ref{sixone})  by integrating
$\exp(iS [\phi(x)]/\hbar)$ over {\it all} field configurations
between
${\cal I}^-$ and
${\cal I}^+$ restricted to $F$ on $\sigma_1$. We can
do the integral first only over fields between ${\cal I}^-$ and
a spacelike surface
$\sigma$, and then push $\sigma$ to ${\cal I}^+$. (See Figure 7.)
The integral over fields between ${\cal I}^-$ and a spacelike surface
$\sigma$ on which a given spatial field configuration $\chi(\bf x)$ is
specified defines the field representative $\langle \chi({\bf x}) |
\Psi(\sigma) \rangle$ of a state 
$|\Psi(\sigma)\rangle$ on $\sigma$. As $\sigma$ is  pushed forward
this integral defines the evolution of $|\Psi(\sigma)\rangle$. 
As Feynman showed, this evolution is unitary between 
infinitesimally separated surfaces.  As a
consequence $|\Psi(\sigma)\rangle$ evolves unitarily up to $\sigma_1$.
There, the restriction of the sum over fields to a set $F$ is equivalent
to the action of the projection on $F$ on $|\Psi(\sigma)\rangle$. 
The state is reduced as in (\ref{twoone b}).  Thus, a state evolving
unitarily or by reduction is recovered from the more general sum-over-histories
formulation in globally hyperbolic spacetimes.

In an evaporating black hole spacetime (Figure 8), the probabilities
for evolution --- for decoherent alternatives on $A$ given an initial
state on $\sigma'$ for instance --- are generally defined four-dimensionally
through (\ref{sixthree}) and (\ref{sixfour}). However, that evolution cannot
be reproduced by the unitary
evolution of a state on spacelike surfaces.
If one attempts to define a state 
 $|\Psi(\sigma)\rangle$ by the procedure described above for
globally hyperbolic spacetimes, one finds that the surface $\sigma$
cannot remain spacelike and be pushed smoothly into the lightly shaded 
region to the future of the
naked singularity.  

The best that can be done is to push the surface $\sigma$ so that
integration is over fields with support in the region bounded by
${\cal I}^-$ and the spacelike surfaces $A$ and $B$ shown in  Figure 8. 
That defines a kind of two component
``state'' with one component on $A$ and the other on $B$. By tracing
products of this object over the degrees of freedom on $B$ and 
normalizing the result, a density
matrix may be constructed that is sufficient for predictions of 
alternatives on $A$.
However, that density matrix does not evolve by anything like an 
equation of the form 
(\ref{threetwo}). Indeed it evolves unitarily as the surface
$A$ is pushed forward in time.  There is no problem with
conservation of energy or charge because of the
general arguments given in \cite{HLM95}. For alternatives
that refer to field configurations in the interior of the spacetime
even this kind of density matrix construction is in general unavailable
and the general expressions (\ref{sixthree}) and (\ref{sixfour}) must
be used.\footnote{It should be noted that a generalized quantum 
theory  formulated four-dimensionally in terms of histories that
extend over the whole of a background spacetime is not necessarily 
causal in the sense that predictions on a spacelike surface are
independent of the background geometry to its future. See
\cite{Har95a,And95,RosZZ} for further discussion.}

Similar statements also could be made  for other spacetimes that are not
globally hyperbolic, such as the spacetimes with closed timelike curves
or spatial topology change
mentioned in Section II, and for quantum gravity itself.

The process of black hole evaporation is thus not in conflict with the
principles of quantum mechanics suitably generally stated.  It is not in
conflict with quantum evolution described four-dimensionally.  It is
only in conflict with the narrow idea that this evolution be reproduced
by evolution of a state vector through a family of spacelike surfaces.

\section{Information --- Where is It?}

A spacetime formulation of quantum mechanics requires a spacetime notion
of information that is also in fully four-dimensional form.  
In this section we describe a notion of the\hfill\break
\vskip .13in
\centerline{\epsfysize=3.00in \epsfbox{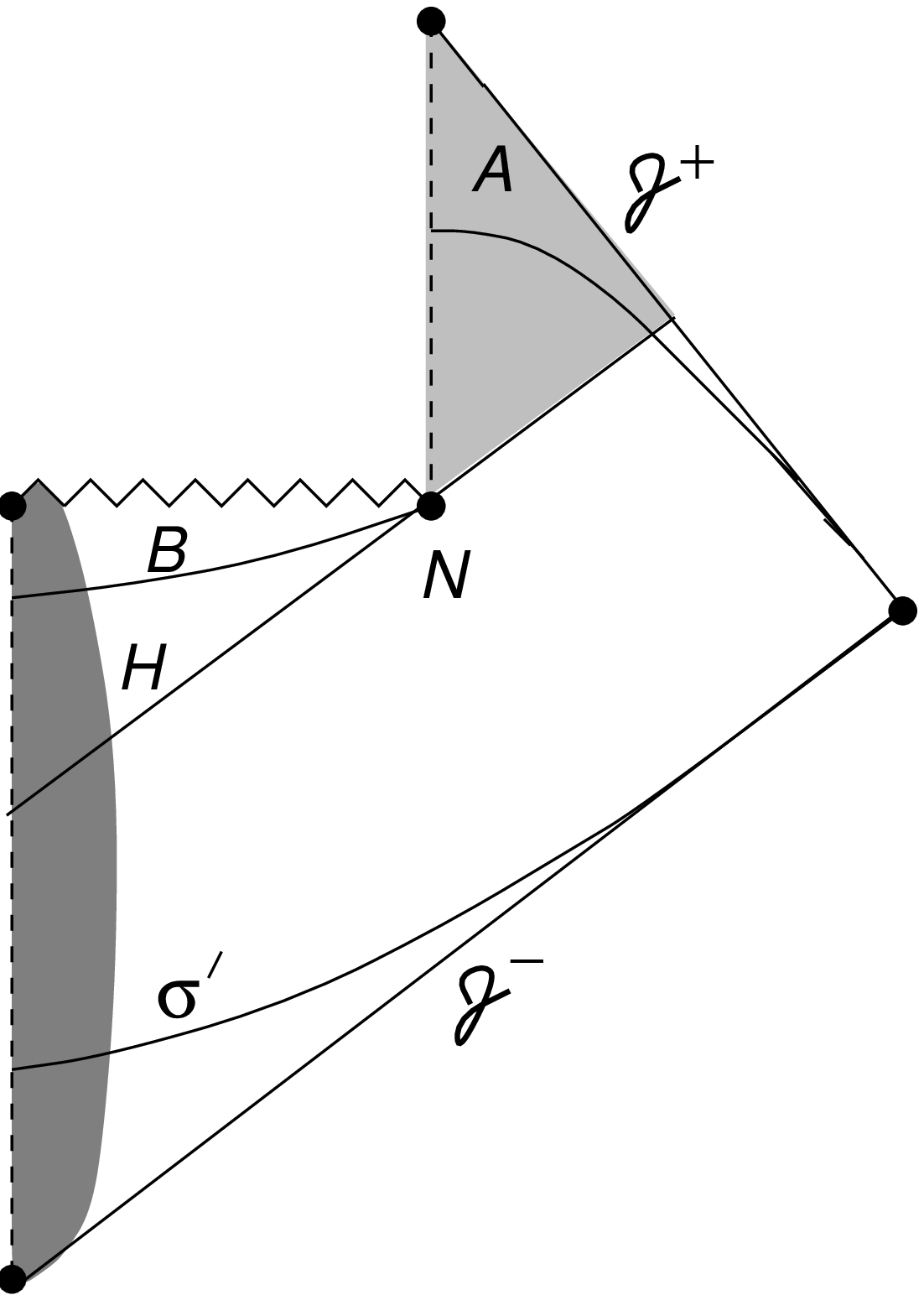}}
\vskip .13 in
\begin{quote}
{{\bf Figure 8:} {\sl Quantum evolution in an evaporating black
hole spacetime can be described four-dimensionally using a
Feynman sum-over-histories. However, that evolution is not 
expressible in $3+1$ terms
by the smooth evolution of a state through spacelike surfaces.

Complete information is not recoverable on surface $A$ because of
correlations (entanglements) between the field on $A$ and the field on a
surface like $B$. Even though information is not necessarily
available on any one spacelike surface, it  is not lost in an
evaporating black hole geometry but it is distributed over
four-dimensional spacetime.}}
\end{quote}
\vskip .13 in
information available in {\it histories} and
not just in alternatives on a single spacelike surface.\footnote{The
particular construction we use is due to Gell-Mann and the author
\cite{Har95a}. There are a number of other ideas, {\it e.g.}~\cite{ILup,GHup}
 with which the same
points about information in a evaporating black-hole spacetime could be
made.} We then apply this to discuss the question of whether information
is lost in an evaporating black-hole spacetime. 

In quantum mechanics, a statistical distribution of states is described
by a density matrix. For the forthcoming discussion it is therefore
necessary to generalize the previous considerations a bit and treat mixed
density matrices $\rho$ as initial conditions for the closed system as
well as pure states $|\Psi\rangle$. To do this it is only necessary to
replace (\ref{fivefive}) with
\begin{equation}
D(\alpha^\prime,\alpha) = Tr\left(C_{\alpha^\prime}\rho
C^\dagger_\alpha\right)\ .
\label{sevenone}
\end{equation}

A generalization of the standard Jaynes construction \cite{Ros83}
gives a natural definition of the missing information in a set of
histories $\{c_\alpha\}$. We begin by defining the
entropy functional on density matrices:

\begin{equation}
{\cal S}(\tilde\rho)  \equiv -Tr\left(\tilde\rho \log \tilde\rho\right)\ .
\label{seventwo}
\end{equation}
With this we define the {\it missing} {\it information}
$S(\{c_\alpha\})$ in a set of histories $\{c_\alpha\}$ as the maximum
of ${\cal S} (\tilde\rho)$ over all density matrices $\tilde\rho$ that
preserve the predictions of the true density matrix $\rho$ for the
decoherence and probabilities of the set of histories $\{c_\alpha\}$. Put differently, we
maximize ${\cal S} (\tilde\rho)$ over $\tilde\rho$ that preserve the
decoherence functional of $\rho$ defined in terms of the corresponding
class operators $\{C_\alpha\}$. 
Thus,
the missing information in a set of histories $\{c_\alpha\}$ is given
explicitly by:
\begin{equation}
S(\{c_\alpha\}) =  \mathop{\rm max}_{\tilde\rho} \Bigl[{\cal S}
(\tilde\rho)\Bigr]_{\{Tr(C_{\alpha^\prime}\tilde\rho
C^\dagger_\alpha) = Tr(C_{\alpha^\prime}\rho
C^\dagger_\alpha)\}}\ .
\label{sevenfour}
\end{equation}

{\it Complete} {\it information}, $S_{\rm compl}$ --- the most one can
have about the initial $\rho$ --- is found in the decoherent set of
histories with the least missing information.
\begin{equation}
S_{\rm compl} = \mathop{\rm min}_{{\rm decoherent} \atop \{c_\alpha\}}
\Bigl[S(\{c_\alpha\})\Bigr]\ .
\label{sevenfive}
\end{equation}
In usual quantum mechanics it is not difficult to show that $S_{\rm
compl}$ defined in this way  is exactly the missing information in the
initial density matrix $\rho$
\begin{equation}
S_{\rm compl} = {\cal S}(\rho) = - Tr(\rho \log \rho)\ .
\label{sevensix}
\end{equation}

Information in a set of histories and complete information are {\it spacetime} notions of information whose construction makes no
reference to states or alternatives on a spacelike surfaces.  Rather the
constructions are four-dimensional
making use of histories.  Thus for
example, with these notions one can capture the idea of information in
entanglements in time as well as information in entanglements in space.

One can find {\it where} information is located in
spacetime by asking what information is available from alternatives
restricted to fields in various spacetime regions.  
For example, alternative values of a field average over a region $R$
refer only to fields inside $R$. The missing information in a
region $R$ is 
\begin{equation}
S(R) = \mathop{\rm min}_{{\rm decohering}\{c_\alpha\} 
\atop{\rm referring\ to}\ R} 
S(\{c_\alpha\})\ .
\label{sevenseven}
\end{equation}
This region could be part of a spacelike surface or could extend in
time; it could be connected or disconnected.  When $R$ is extended over
the whole of the spacetime the missing information is complete.  But it
is an
interesting question what {\it smaller} regions $R$ contain complete
information, $S(R) = S_{\rm compl}$.

When spacetime is globally hyperbolic and quantum dynamics can be
described by states unitarily evolving through a foliating family of
spacelike surfaces, it is a consequence of the definitions
(\ref{sevenseven}) and (\ref{sevenfive}) that complete information is
available on each and every spacelike surface
\begin{equation}
S(\sigma) = S_{\rm compl} = - Tr (\rho \log \rho)\ .
\label{seveneight}
\end{equation}
However, in more general cases, such as the evaporating black hole
spacetimes under consideration in this paper, only incomplete
information may be available on any spacelike surface, and complete
information may be distributed about the spacetime.  We now show why.

There is no reason to expect to recover complete information on
a surface like $A$ in Figure 8. The analysis of the black-hole spacetime without
back reaction (Figure 2) shows that much of the information on the
surface $({\cal I}^+, H)$ consists of correlations, or entanglements,
between alternatives on ${\cal I}^+$ and alternatives on $H$. One should
rather expect complete information to be available in the spacetime
region which is the union of $A$ and a surface like $B$. Even though
this region is not a spacelike surface there are still entanglements
``in time'' between alternatives on $A$ and alternatives on $B$ that
must be considered to completely account for missing information.

The situation is not so very different from that of the ``trousers''
spacetime sketched in Figure 5. Complete information about an initial
state on $\sigma^\prime$ is plausibly not available on surfaces $A$ or
$B$ separately. That is because there will generally be correlations
(entanglements) between alternatives on $A$ and alternatives on 
$B$. Complete information is thus
available in spacetime even if not available on any one spacelike
surface like $A$. The surfaces $A$ and $B$ 
of this example are similar in this respect
to the surfaces $A$ and $B$ of the evaporating black hole spacetime
in Figure 8.

Thus even
though it is not completely available on every spacelike surface like
$A$, information is not lost in evaporating black-hole spacetimes. 
Complete information is distributed about the spacetime. 

\section{Conclusions}

This paper has made five points:

\begin{itemize}
\item For quantum dynamics to be formulated in terms of a state vector
evolving unitarily or by reduction through a family of spacelike
surfaces, spacetime geometry must be fixed and foliable by a
non-singular family of spacelike surfaces.

\item When spacetime geometry is fixed but does not admit a foliating family
of spacelike surfaces or where, as in quantum gravity, spacetime
geometry is not fixed, quantum evolution cannot be defined in terms of
states on spacelike surfaces.

\item But quantum mechanics can be generalized so that it is fully
four-dimensional, spacetime form, free from the need for states on
spacelike surfaces.

\item The complete evaporation of a black hole is not in conflict with
the principles of quantum mechanics stated suitably generally in
four-dimensional form.

\item 
Even though complete information may not be available on every spacelike
surface, four-dimensional information
is not lost in an evaporating black hole spacetime.  
Rather, complete information is distributed over spacetime.
\end{itemize}

We have argued that, whether  or not a quantum theory of gravity exhibits
a unitary $S$-matrix between pre-collapse and evaporated states, some
generalization of quantum mechanics will be necessary when spacetime
geometry is not fixed.  Black hole evaporation is not in conflict
with the principles of quantum mechanics suitably generally stated.
Whatever the outcome of the evaporation process generalized quantum
theory is ready to describe them.

\section{Epilogue}

When I was a much younger scientist working on the physics of neutron
stars, I remember reporting to Chandra how I had been criticized after a
colloquium for
daring to extrapolate the theory of matter in its ground state to the
densities of nuclear matter and beyond. He emphatically advised me to
pay no attention, as if worried that I might, saying people such as my critics
were never right.  He might, I imagine, have been thinking of his own
extrapolation of the properties of stellar matter to the densities of
white dwarfs.  In this article I have been engaged in extrapolating the
principles of quantum mechanics to the domains of quantum gravity and
black holes.  One hopes Chandra would have approved.

\acknowledgments

This work was supported in part by NSF grants PHY95-07065 and
PHY94-07194.

\end{document}